\documentclass{article}
\usepackage{graphicx}
\usepackage{amsmath}
\usepackage{amssymb}
\usepackage{amscd}
\usepackage{amsfonts}

 \evensidemargin0in \oddsidemargin0in \topmargin10pt
\begin{document}

\title{Entangled state for constructing generalized phase space
representation and its statistical behavior}
\author{$^{1}${\small Li-yun Hu and }$^{1,2}${\small Hong-yi Fan} \\
$^{1}${\small Department of Physics, Shanghai Jiao Tong University, }\\
{\small Shanghai, 200030, China}\\
$^{2}${\small Department of Material Science and Engineering, University of}%
\\
{\small Science and Technology of China, Hefei, Anhui 230026, China}}
\maketitle

\begin{abstract}
{\small Based on the conception of quantum entanglement of
Einstein-Podolsky-Rosen} {\small we construct generalized phase space
representation associated with the entangled state }$\left\vert \Gamma
\right\rangle _{e}${\small , which is endowed with definite physical
meaning. The set of states make up a complete and non-orthogonal
representation. The Weyl ordered form of }$\left\vert \Gamma \right\rangle
_{ee}\left\langle \Gamma \right\vert ${\small \ is derived which clearly
exhibit the statistical behavior of marginal distribution of }$\left\vert
\Gamma \right\rangle _{ee}\left\langle \Gamma \right\vert .${\small \ The
minimum uncertainty relation obeyed by }$\left\vert \Gamma \right\rangle
_{e} ${\small \ is also demonstrated.}
\end{abstract}

\section{Introduction}

Phase space formalism of quantum mechanics has many applications in quantum
statistics, quantum optics and quantum information theory. It began with
Wigner's celebrated paper in 1932 \cite{r1}. Among many kinds of
pseudo-probability distribution functions, Wigner function $W(q,p)$ of a
quantum state (pure or mixed states) is the most popularly used, since in
phase space it exhibits$\ $two marginal distribution as the following way
\cite{r2},
\begin{equation}
\mathrm{P}(p)=\int_{-\infty }^{\infty }W(q,p)dq,\;\mathrm{P}\left( q\right)
=\int_{-\infty }^{\infty }W(q,p)dp,  \label{1}
\end{equation}%
where $\mathrm{P}\left( q\right) \;\left[ \mathrm{P}(p)\right] \;$is
proportional to the probability for finding the particle at $q$ [at $p$ in
momentum space]. Besides, the Wigner operator also serves as an integral
kernel of the Weyl rule \cite{r15,r16} which is a quantization scheme
connecting classical functions of ($q,p$) with their quantum correspondence
operators of ($Q,P$). The single-mode Wigner operator in the coordinate
representation is
\begin{equation}
\Delta (q,p)=\frac{1}{2\pi }\int_{-\infty }^{\infty }\left\vert q-\frac{v}{2}%
\right\rangle \left\langle q+\frac{v}{2}\right\vert e^{-ipv}dv,  \label{2}
\end{equation}%
where $\left\vert q\right\rangle $ is the coordinate eigenvector, $%
Q\left\vert q\right\rangle =q\left\vert q\right\rangle ,$
\begin{equation}
\left\vert q\right\rangle =\pi ^{-1/4}\exp \left[ -\frac{q^{2}}{2}+\sqrt{2}%
qa^{\dagger }-\frac{1}{2}a^{\dagger 2}\right] \left\vert 0\right\rangle .
\end{equation}%
Using the normally ordered form of vacuum projector $\left\vert
0\right\rangle \left\langle 0\right\vert =\colon \exp [-a^{\dagger }a]\colon
,$ where $\colon \colon $ denotes normal ordering, and the technique of
integration within an ordered product (IWOP) of operators \cite{r10,r11}$,$
the integration in Eq. (\ref{2}) can be performed, leading to the explicit
operator
\begin{equation}
\Delta (q,p)=\frac{1}{\pi }\colon e^{-(q-Q)^{2}-(p-P)^{2}}\colon =\frac{1}{%
\pi }\colon e^{-2(a^{\dagger }-\alpha ^{\ast })(a-\alpha )}\colon \equiv
\Delta (\bar{\alpha},\bar{\alpha}^{\ast }),\;\;  \label{3}
\end{equation}%
where $Q$ and $P$ are related to the Bose creation and annihilation
operators $\left( a^{\dagger },a\right) $ by $Q=(a+a^{\dagger })/\sqrt{2}$
and $P=(a-a^{\dagger })/(i\sqrt{2}),$ respectively, $[a,a^{\dagger }]=1,$ $%
\bar{\alpha}=\left( q+ip\right) /\sqrt{2}.$ Obviously, $\int_{-\infty
}^{\infty }\Delta (q,p)dp=\frac{1}{\sqrt{\pi }}\colon e^{-(q-Q)^{2}}\colon
=\left\vert q\right\rangle \left\langle q\right\vert ,$ $\int_{-\infty
}^{\infty }\Delta (q,p)dq=\frac{1}{\sqrt{\pi }}\colon e^{-(p-P)^{2}}\colon
=\left\vert p\right\rangle \left\langle p\right\vert ,$ where $\left\vert
p\right\rangle $ is the momentum eigenvector. This formalism helps us to
understand the Wigner function's role better\cite{r2}\cite{r3}-\cite{r6}.

In two-mode case when two systems are prepared in an entangled state,
measuring one of the two canonically conjugate variables on one system, the
value for a physical variable in the another system may be inferred with
certainty, this is the quantum entanglement. Because entanglement is now
widely used in quantum information and quantum computation, it has been paid
much attention by physicists \cite{r7}. The original idea of quantum
entanglement began with Einstein-Podolsky-Rosen's observation in EPR's
treatment \cite{r8}, who noticed that two particles' relative coordinate $%
Q_{1}-Q_{2}$ and total momentum $P_{1}+P_{2}$ can be simultaneously
measured, (also their conjugate variables $\left[ Q_{1}+Q_{2},P_{1}-P_{2}%
\right] =0),$ therefore, the corresponding Wigner function should be such
that its two marginal distributions are respectively proportional to the
probability for finding the two particles which possess certain total
momentum value $\left[ \text{relative momentum value}\right] $ and
simultaneously relative\ position value $\left[ \text{center-of-mass
position value}\right] $ (see also Eqs. (\ref{13}) and (\ref{16})). The
investigation of Wigner functions for entangled states is not only just for
the convenience of some calculations, but also for revealing the intrinsic
entanglement property inherent to some physical systems. In Ref. \cite{r21}
by virtue of the well-behaved properties of the entangled state
representation $\left\langle \eta \right\vert $,
\begin{equation}
\left\vert \eta \right\rangle =\exp \left\{ -\frac{1}{2}\left\vert \eta
\right\vert ^{2}+\eta a_{1}^{\dagger }-\eta ^{\ast }a_{2}^{\dagger
}+a_{1}^{\dagger }a_{2}^{\dagger }\right\} \left\vert 00\right\rangle
,\;\eta =\eta _{1}+i\eta _{2},  \label{8}
\end{equation}%
we have successfully established the so-called entangled Wigner operator for
correlated two-body systems$\ $\cite{r21},
\begin{equation}
\Delta _{w}\left( \rho ,\varsigma \right) =\int \frac{d^{2}\eta }{\pi ^{3}}%
\left\vert \rho -\eta \right\rangle \left\langle \rho +\eta \right\vert \exp
(\eta \varsigma ^{\ast }-\eta ^{\ast }\varsigma ),  \label{4}
\end{equation}%
$\left\vert \eta \right\rangle $ is the common eigenvector of $Q_{1}-Q_{2}$
and $P_{1}+P_{2}$ \cite{r24}, which obeys the eigenvector equations
\begin{equation}
(Q_{1}-Q_{2})\left\vert \eta \right\rangle =\sqrt{2}\eta _{1}\left\vert \eta
\right\rangle ,\;\;(P_{1}+P_{2})\left\vert \eta \right\rangle =\sqrt{2}\eta
_{2}\left\vert \eta \right\rangle .  \label{6}
\end{equation}%
Using the IWOP technique we have shown in\textbf{\ \cite{r21} }that
$\Delta _{w}\left( \rho ,\varsigma \right) $ is just the product of
two independent single-mode Wigner operators $\Delta _{w}\left( \rho
,\varsigma \right)
=\Delta (\bar{\alpha},\bar{\alpha}^{\ast })\Delta (\bar{\beta},\bar{\beta}%
^{\ast })$ provided we take
\begin{equation}
\varsigma =\bar{\alpha}+\bar{\beta}^{\ast },\;\rho =\bar{\alpha}-\bar{\beta}%
^{\ast },\text{\ \ }\bar{\alpha}=\frac{q_{1}+ip_{1}}{\sqrt{2}},\text{ }\bar{%
\beta}=\frac{q_{2}+ip_{2}}{\sqrt{2}}.  \label{12}
\end{equation}%
Performing the integration of $\Delta _{w}\left( \rho ,\varsigma \right) $
over $d^{2}\varsigma $ leads to the projection operator of the entangled
state $\left\vert \eta \right\rangle $
\begin{equation}
\int d^{2}\varsigma \Delta _{w}(\rho ,\varsigma )=\frac{1}{\pi }\left\vert
\eta \right\rangle \left\langle \eta \right\vert |_{\eta =\rho },\text{ }
\label{13}
\end{equation}%
and the marginal distribution in ($\eta _{1},\eta _{2}$) phase space is $%
\left\langle \psi \right\vert \int d^{2}\varsigma \Delta _{w}(\rho
,\varsigma )\left\vert \psi \right\rangle =\frac{1}{\pi }|\psi (\eta
)|^{2}|_{\eta =\rho }.$(in reference to (\ref{6})). Similarly, we can
introduce the common eigenvector of$\;Q_{1}+Q_{2}$ and $P_{1}-P_{2}$ \cite%
{r24} (the conjugate state of $\left\vert \eta \right\rangle )$
\begin{equation}
\left\vert \xi \right\rangle =\exp \left\{ -\frac{1}{2}\left\vert \xi
\right\vert ^{2}+\xi a_{1}^{\dagger }+\xi ^{\ast }a_{2}^{\dagger
}-a_{1}^{\dagger }a_{2}^{\dagger }\right\} \left\vert 00\right\rangle
,\;\;\xi =\xi _{1}+i\xi _{2},  \label{14}
\end{equation}%
which obeys another pair of eigenvector equations
\begin{equation}
\left( Q_{1}+Q_{2}\right) \left\vert \xi \right\rangle =\sqrt{2}\xi
_{1}\left\vert \xi \right\rangle ,\;\;\left( P_{1}-P_{2}\right) \left\vert
\xi \right\rangle =\sqrt{2}\xi _{2}\left\vert \xi \right\rangle .
\label{15e}
\end{equation}%
Performing the integration of $\Delta _{w}\left( \rho ,\varsigma \right) $
over $d^{2}\rho $ yields
\begin{equation}
\int d^{2}\rho \Delta _{w}(\rho ,\varsigma )=\frac{1}{\pi }\left\vert \xi
\right\rangle \left\langle \xi \right\vert |_{\xi =\varsigma
},\;\left\langle \psi \right\vert \int d^{2}\rho \Delta _{w}(\rho ,\varsigma
)\left\vert \psi \right\rangle =\frac{1}{\pi }|\psi (\xi )|^{2}|_{\xi
=\varsigma }.  \label{16}
\end{equation}%
The introduction of the entangled Wigner operator also brings much
convenience for calculating the Wigner function of some entangled states.

Working in the $\left\vert \eta \right\rangle or%
\left\vert \xi \right\rangle $ representation one can %
interrelate some physical systems. For example, the
Einstein-Podolsky-Rosen arrangement relies on free propagation of
quantum-coupled particles (described by $\left\vert \eta
\right\rangle $ or $\left\vert \xi \right\rangle )$, on the other
hand, the two-mode squeezed state dealing with oscillators which are
bound systems, these two seemingly very different physical systems
can be interrelated by constructing the following
ket-bra integration in terms of $\left\vert \eta \right\rangle ,$%
\begin{equation}
\int \frac{d^{2}\eta }{\pi \mu }|\frac{\eta }{\mu }\rangle \langle \eta
|=\exp \left[ \left( a_{1}^{\dagger }a_{2}^{\dagger }-a_{1}a_{2}\right) \ln
\mu \right] ,  \label{16a}
\end{equation}%
the right hand-side of (\ref{16a}) is just the two-mode squeezing
operator \cite{r3,24a}. In the following we shall employ\textbf{\ }the $%
\left\vert \eta \right\rangle $ state to formulating generalized phase space
representation and then study its statistical behaviors, i.e. enlightened by
Eqs. (\ref{6}) and (\ref{15e}) we construct generalized phase space
representation characteristic of the properties under the action of $%
Q_{1}-Q_{2}$ and $P_{1}+P_{2}$, associated with a two-mode state vector $%
\left\vert \Gamma \right\rangle _{e}$,
\begin{eqnarray}
_{e}\left\langle \Gamma \right\vert \frac{Q_{1}-Q_{2}}{\sqrt{2}} &=&\left(
\alpha \sigma _{1}+i\beta \frac{\partial }{\partial \tau _{2}}\right) \text{
}_{e}\left\langle \Gamma \right\vert ,  \notag \\
_{e}\left\langle \Gamma \right\vert \frac{P_{1}-P_{2}}{\sqrt{2}} &=&\left(
\gamma \tau _{2}+i\delta \frac{\partial }{\partial \sigma _{1}}\right) \text{
}_{e}\left\langle \Gamma \right\vert ,  \label{e5}
\end{eqnarray}%
where the subscript \textquotedblleft $e$\textquotedblright\ implies the
entanglement, $\alpha ,\beta ,\gamma $ and $\delta $ are all real
parameters, satisfying%
\begin{equation}
\beta \gamma -\alpha \delta =1,  \label{e3}
\end{equation}%
and $\left[ Q_{1}-Q_{2},P_{1}-P_{2}\right] =2i.$ Simultaneously, under the
action of the center-of-mass operator and the relative momentum operator,
the state $_{e}\left\langle \Gamma \right\vert $ behaves
\begin{eqnarray}
_{e}\left\langle \Gamma \right\vert \frac{Q_{1}+Q_{2}}{\sqrt{2}} &=&\left(
\gamma \tau _{1}-i\delta \frac{\partial }{\partial \sigma _{2}}\right) \text{
}_{e}\left\langle \Gamma \right\vert ,  \notag \\
_{e}\left\langle \Gamma \right\vert \frac{P_{1}+P_{2}}{\sqrt{2}} &=&\left(
\alpha \sigma _{2}-i\beta \frac{\partial }{\partial \tau _{1}}\right) \text{
}_{e}\left\langle \Gamma \right\vert .  \label{e6}
\end{eqnarray}

The present paper is arranged as follows. In Sec. 2 using the newly
developed bipartite entangled state representation $|\eta \rangle $ of
continuum variables we shall derive the concrete form of entangled state $%
\left\vert \Gamma \right\rangle _{e}$ in two-mode Fock space and
then analyze its properties, in so doing the phase space theory can
be developed to the entangled case. In Sec.\ 3, the completeness
relation and non-orthonormal property of $\left\vert \Gamma
\right\rangle _{e}$\ are proved. In Sec. 4. the Weyl ordered form of
$\left\vert \Gamma \right\rangle _{ee}\left\langle \Gamma
\right\vert $ is derived, which yields the classical correspondence
of $\left\vert \Gamma \right\rangle _{ee}\left\langle \Gamma
\right\vert $. In Sec. 5 we examine marginal distributions of the
operator $\left\vert \Gamma \right\rangle _{ee}\left\langle \Gamma
\right\vert $ by using the properties of the entangled state $|\eta
\rangle $ and its conjugate state $\left\vert \xi \right\rangle $.
The uncertainty relation of coordinate and momentum quadratures in
$\left\vert \Gamma \right\rangle _{e}$ and the Wigner function of
$\left\vert \Gamma \right\rangle _{e}$ are calculated in sections 6
and 7, respectively.

\section{The state $\left\vert \Gamma \right\rangle _{e}$ in two-mode Fock
space}

We find that the explicit form of the state $\left\vert \Gamma \right\rangle
_{e}$ in two-mode Fock space is (see the Appendix),

\begin{equation}
\left\vert \Gamma \right\rangle _{e}\equiv 2\sqrt{-\alpha \beta \gamma
\delta }\exp \left[ \frac{\alpha \left\vert \sigma \right\vert ^{2}}{2\delta
}-\frac{\gamma \left\vert \tau \right\vert ^{2}}{2\beta }+\left( \alpha
\sigma +\gamma \tau \right) a_{1}^{\dagger }+\left( \gamma \tau ^{\ast
}-\alpha \sigma ^{\ast }\right) a_{2}^{\dagger }-\left( \beta \gamma +\alpha
\delta \right) a_{1}^{\dagger }a_{2}^{\dagger }\right] \left\vert
00\right\rangle ,  \label{20}
\end{equation}%
where $\sigma =\sigma _{1}+i\sigma _{2},$ $\tau =\tau _{1}+i\tau _{2}$; real
numbers ($\alpha ,\beta ,\gamma $ and $\delta $) satisfy the relation Eq.(%
\ref{e3}); $(a_{i},a_{i}^{\dag }),$ $i=1,2,$ are the two-mode Bose
annihilation and creation operators obeying $\left[ a_{i},a_{j}^{\dagger }%
\right] =\delta _{ij}$. To satisfy the square integrable condition for wave
function in phase space $\left\vert \Gamma \right\rangle _{e},$ $\frac{%
\alpha }{\delta }<0$ and $\frac{\gamma }{\beta }>0$ are demanded. In order
to certify that Eq.(\ref{20}) really obeys Eqs. (\ref{e5}) and (\ref{e6}) we
operate $a_{i}$ on $\left\vert \Gamma \right\rangle _{e},$
\begin{eqnarray}
a_{1}\left\vert \Gamma \right\rangle _{e} &=&\left[ \left( \alpha \sigma
+\gamma \tau \right) -\left( \beta \gamma +\alpha \delta \right)
a_{2}^{\dagger }\right] \left\vert \Gamma \right\rangle _{e},  \notag \\
a_{2}\left\vert \Gamma \right\rangle _{e} &=&\left[ \left( \gamma \tau
^{\ast }-\alpha \sigma ^{\ast }\right) -\left( \beta \gamma +\alpha \delta
\right) a_{1}^{\dagger }\right] \left\vert \Gamma \right\rangle _{e}.
\label{23}
\end{eqnarray}%
Then noting the relation between $Q_{i},P_{j}$ and $a_{i},a_{j}^{\dag }$,%
\begin{equation}
Q_{i}=(a_{i}+a_{i}^{\dag })/\sqrt{2},\ P_{i}=(a_{i}-a_{i}^{\dag })/(\sqrt{2}%
\mathtt{i}),  \label{e24}
\end{equation}%
and Eq.(\ref{e3}) as well as%
\begin{eqnarray}
\frac{\partial }{\partial \sigma }\left. _{e}\left\langle \Gamma \right\vert
\right. &=&\left. _{e}\left\langle \Gamma \right\vert \left( \frac{\alpha
\sigma ^{\ast }}{2\delta }-\alpha a_{2}\right) \right. ,\text{ }\frac{%
\partial }{\partial \sigma ^{\ast }}\left. _{e}\left\langle \Gamma
\right\vert \right. =\left. _{e}\left\langle \Gamma \right\vert \left( \frac{%
\alpha \sigma }{2\delta }+\alpha a_{1}\right) \right. ,  \notag \\
\frac{\partial }{\partial \tau }\left. _{e}\left\langle \Gamma \right\vert
\right. &=&\left. _{e}\left\langle \Gamma \right\vert \left( -\frac{\gamma
\tau ^{\ast }}{2\beta }+\gamma a_{2}\right) \right. ,\text{ }\frac{\partial
}{\partial \tau ^{\ast }}\left. _{e}\left\langle \Gamma \right\vert \right.
=\left. _{e}\left\langle \Gamma \right\vert \left( -\frac{\gamma \tau }{%
2\beta }+\gamma a_{1}\right) \right. ,  \label{24}
\end{eqnarray}%
we see, for example,%
\begin{eqnarray}
_{e}\left\langle \Gamma \right\vert \frac{Q_{1}+Q_{2}}{\sqrt{2}} &=&\left.
_{e}\left\langle \Gamma \right\vert \right. \left[ -\delta \left( \alpha
a_{1}+\alpha a_{2}\right) -i\alpha \sigma _{2}+\gamma \tau _{1}\right]
\notag \\
&=&\left[ \gamma \tau _{1}+\delta \left( \frac{\partial }{\partial \sigma }-%
\frac{\partial }{\partial \sigma ^{\ast }}\right) \right] \left.
_{e}\left\langle \Gamma \right\vert \right.  \notag \\
&=&\left( \gamma \tau _{1}-i\delta \frac{\partial }{\partial \sigma _{2}}%
\right) \left. _{e}\left\langle \Gamma \right\vert \right. ,  \label{25}
\end{eqnarray}%
which is the first equation in Eq.(\ref{e6}). In a similar way, $%
_{e}\left\langle \Gamma \right\vert $ satisfying the other equations in Eqs.(%
\ref{e5}) and (\ref{e6}) can be checked. Using Eqs.(\ref{e5}), (\ref{e6})
and noticing the quantum commutator%
\begin{equation}
\left[ \frac{Q_{1}\pm Q_{2}}{\sqrt{2}},\frac{P_{1}\pm P_{2}}{\sqrt{2}}\right]
=i,  \label{29}
\end{equation}%
in the phase space representation, we have

\begin{equation}
_{e}\left\langle \Gamma \right\vert \left[ \frac{Q_{1}\pm Q_{2}}{\sqrt{2}},%
\frac{P_{1}\pm P_{2}}{\sqrt{2}}\right] =i\left( \beta \gamma -\alpha \delta
\right) _{e}\left\langle \Gamma \right\vert ,  \label{30}
\end{equation}%
which results in the condition shown in Eq.(\ref{e3}).

\section{The properties of $\left\vert \Gamma \right\rangle _{e}$}

\subsection{The completeness relation of $\left\vert \Gamma \right\rangle
_{e}$}

Next we prove the completeness relation of Eq.(\ref{20}). Using the normally
ordered vacuum projector
\begin{equation}
\left\vert 00\right\rangle \left\langle 00\right\vert =\colon \exp \left(
-a_{1}^{\dagger }a_{1}-a_{2}^{\dagger }a_{2}\right) \colon ,  \label{31}
\end{equation}%
where $\colon \colon $ denotes the normal product, which means all the
bosonic creation operators are standing on the left of annihilation
operators in a monomial of $a^{\dagger }$ and $a$ \cite{r9}. It should be
emphasized that a normally ordered product of operators can be integrated
with respect to $c$-numbers provided the integration is convergent. Then we
can use Eq.(\ref{20}) and the IWOP technique to perform the following
integration%
\begin{eqnarray}
&&\frac{1}{\beta ^{2}\delta ^{2}}\int \frac{d^{2}\sigma d^{2}\tau }{4\pi ^{2}%
}\left\vert \Gamma \right\rangle _{ee}\left\langle \Gamma \right\vert  \notag
\\
&=&-\frac{\alpha \gamma }{\beta \delta }\int \frac{d^{2}\sigma d^{2}\tau }{%
\pi ^{2}}\colon \exp \left[ \frac{\alpha \left\vert \sigma \right\vert ^{2}}{%
\delta }+\sigma \alpha \left( a_{1}^{\dagger }-a_{2}\right) +\sigma ^{\ast
}\alpha \left( a_{1}-a_{2}^{\dagger }\right) -a_{1}^{\dagger }a_{1}\right.
\notag \\
&&\left. -\frac{\gamma \left\vert \tau \right\vert ^{2}}{\beta }+\tau \gamma
\left( a_{1}^{\dagger }+a_{2}\right) +\tau ^{\ast }\gamma \left(
a_{2}^{\dagger }+a_{1}\right) -\left( \beta \gamma +\alpha \delta \right)
\left( a_{1}^{\dagger }a_{2}^{\dagger }+a_{1}a_{2}\right) -a_{2}^{\dagger
}a_{2}\right] \colon  \notag \\
&=&-\frac{\alpha \gamma }{\beta \delta }\int \frac{d^{2}\sigma d^{2}\tau }{%
\pi ^{2}}\colon \exp \left\{ \frac{\alpha }{\delta }\left[ \sigma +\delta
\left( a_{1}-a_{2}^{\dagger }\right) \right] \left[ \sigma ^{\ast }+\delta
\left( a_{1}^{\dagger }-a_{2}\right) \right] \right.  \notag \\
&&\left. -\frac{\gamma }{\beta }\left[ \tau -\beta \left( a_{2}^{\dagger
}+a_{1}\right) \right] \left[ \tau ^{\ast }-\beta \left( a_{1}^{\dagger
}+a_{2}\right) \right] \right\} \colon  \notag \\
&=&\colon \exp \left[ -\left( a_{1}^{\dagger }a_{1}+a_{2}^{\dagger
}a_{2}\right) \left( \alpha \delta -\beta \gamma +1\right) \right] \colon =1,
\label{32}
\end{eqnarray}%
where we have used the integral formula \cite{r12}
\begin{equation}
\int \frac{d^{2}\beta }{\pi }\exp \left[ \varsigma \left\vert \beta
\right\vert ^{2}+\xi \beta +\eta \beta ^{\ast }\right] =-\frac{1}{\varsigma }%
\exp \left[ -\frac{\xi \eta }{\varsigma }\right] ,\text{ Re}\varsigma <0.
\label{33}
\end{equation}%
Thus $\left\vert \Gamma \right\rangle _{e}$ is capable of making up a new
quantum mechanical representation.

\subsection{The non-orthonormal property of $\left\vert \Gamma \right\rangle
_{e}$}

Noticing the overlap relation
\begin{eqnarray}
_{e}\left\langle \Gamma \right\vert \left. z_{1},z_{2}\right\rangle &=&2%
\sqrt{-\alpha \beta \gamma \delta }\exp \left[ -\frac{\left\vert
z_{1}\right\vert ^{2}}{2}+\frac{\alpha \left\vert \sigma \right\vert ^{2}}{%
2\delta }-\frac{\gamma \left\vert \tau \right\vert ^{2}}{2\beta }+\left(
\alpha \sigma ^{\ast }+\gamma \tau ^{\ast }\right) z_{1}\right]  \notag \\
&&\times \exp \left[ -\frac{\left\vert z_{2}\right\vert ^{2}}{2}+\left(
\gamma \tau -\alpha \sigma \right) z_{2}-\left( \beta \gamma +\alpha \delta
\right) z_{1}z_{2}\right] ,  \label{34}
\end{eqnarray}%
where $\left\vert z\right\rangle =\exp \left( -\left\vert z\right\vert
^{2}/2+za^{\dagger }\right) \left\vert 0\right\rangle $ is the coherent
state \cite{r13,r14} and using the over-completeness relation of coherent
states $\int \frac{d^{2}z_{1}d^{2}z_{2}}{\pi ^{2}}\left\vert
z_{1},z_{2}\right\rangle \left\langle z_{1},z_{2}\right\vert =1,$ we can
derive the inner-product $_{e}\left\langle \Gamma \right\vert \left. \Gamma
^{\prime }\right\rangle _{e}$, ($\left\vert \Gamma ^{\prime }\right\rangle
_{e}$ has the same $\beta ,\gamma ,\alpha $ and $\delta $ with $\left\vert
\Gamma \right\rangle _{e}$)$,$

\begin{eqnarray}
_{e}\left\langle \Gamma \right\vert \left. \Gamma ^{\prime }\right\rangle
_{e} &=&\int \frac{d^{2}z_{1}d^{2}z_{2}}{\pi ^{2}}\left. _{e}\left\langle
\Gamma \right. \left\vert z_{1},z_{2}\right\rangle \right. \left\langle
z_{1},z_{2}\right\vert \left. \Gamma ^{\prime }\right\rangle _{e}  \notag \\
&=&-4\alpha \beta \gamma \delta \int \frac{d^{2}z_{1}d^{2}z_{2}}{\pi ^{2}}%
\exp \left[ -\left\vert z_{1}\right\vert ^{2}+\left( \alpha \sigma ^{\ast
}+\gamma \tau ^{\ast }\right) z_{1}+\left( \alpha \sigma ^{\prime }+\gamma
\tau ^{\prime }\right) z_{1}^{\ast }\right]  \notag \\
&&-\left\vert z_{2}\right\vert ^{2}+\left( \gamma \tau -\alpha \sigma
\right) z_{2}+\left( \gamma \tau ^{\prime \ast }-\alpha \sigma ^{\prime \ast
}\right) z_{2}^{\ast }-\left( \beta \gamma +\alpha \delta \right) z_{1}z_{2}
\notag \\
&&\left. +\frac{\alpha }{2\delta }\left( \left\vert \sigma \right\vert
^{2}+\left\vert \sigma ^{\prime }\right\vert ^{2}\right) -\frac{\gamma }{%
2\beta }\left( \left\vert \tau \right\vert ^{2}+\left\vert \tau ^{\prime
}\right\vert ^{2}\right) -\left( \beta \gamma +\alpha \delta \right)
z_{1}^{\ast }z_{2}^{\ast }\right] .  \label{35}
\end{eqnarray}%
With the aid of the integral formula Eq.(\ref{33}), we perform the integral
over $d^{2}z_{1}d^{2}z_{2}$ in Eq.(\ref{35}) and finally obtain%
\begin{eqnarray}
_{e}\left\langle \Gamma \right\vert \left. \Gamma ^{\prime }\right\rangle
_{e} &=&\exp \left[ \frac{\alpha }{4\beta \gamma \delta }\left\vert \sigma
-\sigma ^{\prime }\right\vert ^{2}-\frac{1}{4\beta \delta }\left( \tau
^{\prime }\sigma ^{\ast }-\sigma \tau ^{\prime \ast }+\sigma ^{\prime }\tau
^{\ast }-\tau \sigma ^{\prime \ast }\right) \right.  \notag \\
&&\left. +\frac{\gamma }{4\alpha \beta \delta }\left\vert \tau -\tau
^{\prime }\right\vert ^{2}-\frac{\left( \beta \gamma +\alpha \delta \right)
}{4\beta \delta }\left( \tau ^{\prime }\sigma ^{\prime \ast }-\sigma
^{\prime }\tau ^{\prime \ast }+\sigma \tau ^{\ast }-\tau \sigma ^{\ast
}\right) \right] .  \label{36}
\end{eqnarray}%
From Eq.(\ref{36}) one can see that $_{e}\left\langle \Gamma \right\vert
\left. \Gamma ^{\prime }\right\rangle _{e}$ is non-orthogonal, only when $%
\sigma =\sigma ^{\prime }$ and $\tau =\tau ^{\prime }$, $_{e}\left\langle
\Gamma \right\vert \left. \Gamma \right\rangle _{e}=1.$

\section{The Weyl ordered form of $\left\vert \Gamma \right\rangle
_{ee}\left\langle \Gamma \right\vert $}

For a density operator $\rho $ of bipartite system, we can convert it into
its Weyl ordered form \cite{r15,r16,r17} by using the formula
\begin{equation}
\rho =4\int \frac{d^{2}z_{1}d^{2}z_{2}}{\pi
^{2}}\genfrac{}{}{0pt}{}{:}{:}\left\langle -z_{1},-z_{2}\right\vert
\rho \left\vert z_{1},z_{2}\right\rangle \exp \left[
2\sum_{i=1}^{2}\left( a_{i}^{\dagger }a_{i}+a_{i}z_{i}^{\ast
}-z_{i}a_{i}^{\dagger }\right) \right] \genfrac{}{}{0pt}{}{:}{:},
\label{43}
\end{equation}%
where the symbol $\genfrac{}{}{0pt}{}{:}{:}\genfrac{}{}{0pt}{}{:}{:}$ denotes the Weyl ordering, $%
\left\vert z_{i}\right\rangle $ is the coherent state, $\left\langle
-z_{i}\right\vert \left. z_{i}\right\rangle =\exp \{-2\left\vert
z_{i}\right\vert ^{2}\}$. Note that the order of Bose operators $a_{i}$ and $%
a_{i}^{\dagger }$ within a Weyl ordered product can be permuted. That is to
say, even though $\left[ a,a^{\dagger }\right] =1$, we can have $\genfrac{}{}{0pt}{}{:}{:}%
aa^{\dagger
}\genfrac{}{}{0pt}{}{:}{:}=\genfrac{}{}{0pt}{}{:}{:}a^{\dagger
}a\genfrac{}{}{0pt}{}{:}{:}.$ Substituting Eq.(\ref{20}) into
Eq.(\ref{43}) and performing the integration by virtue of the
technique of integration within a Weyl ordered product
(IWWOP) of operators \cite{r18}, we finally obtain%
\begin{eqnarray}
\left\vert \Gamma \right\rangle _{ee}\left\langle \Gamma \right\vert
&=&-16\alpha \beta \gamma \delta \int \frac{d^{2}z_{1}d^{2}z_{2}}{\pi ^{2}}%
\genfrac{}{}{0pt}{}{:}{:}\exp \left[ -\left\vert z_{1}\right\vert
^{2}+\left( \sigma ^{\ast }\alpha +\tau ^{\ast }\gamma
-2a_{1}^{\dagger }\right) z_{1}+\left(
2a_{1}-\sigma \alpha -\tau \gamma \right) z_{1}^{\ast }\right.   \notag \\
&&-\left\vert z_{2}\right\vert ^{2}+\left( \tau \gamma -\sigma \alpha
-2a_{2}^{\dagger }\right) z_{2}+\left( 2a_{2}-\tau ^{\ast }\gamma +\sigma
^{\ast }\alpha \right) z_{2}^{\ast }  \notag \\
&&\left. -\left( \beta \gamma +\alpha \delta \right) \left( z_{1}^{\ast
}z_{2}^{\ast }+z_{1}z_{2}\right) +\frac{\alpha \left\vert \sigma \right\vert
^{2}}{\delta }-\frac{\gamma \left\vert \tau \right\vert ^{2}}{\beta }%
+2a_{1}^{\dagger }a_{1}+2a_{2}^{\dagger }a_{2}\right] \genfrac{}{}{0pt}{}{:}{:}  \notag \\
&=&4\genfrac{}{}{0pt}{}{:}{:}\exp \left\{ \frac{\alpha \delta
}{\beta \gamma }\left( \frac{\sigma }{\delta }+\left(
a_{1}-a_{2}^{\dagger }\right) \right) \left( \frac{\sigma ^{\ast
}}{\delta }+\left( a_{1}^{\dagger }-a_{2}\right) \right)
\right.   \notag \\
&&\left. +\frac{\gamma \beta }{\alpha \delta }\left( \frac{\tau }{\beta }%
-\left( a_{1}+a_{2}^{\dagger }\right) \right) \left( \frac{\tau ^{\ast }}{%
\beta }-\left( a_{2}+a_{1}^{\dagger }\right) \right) \right\}
\genfrac{}{}{0pt}{}{:}{:}, \label{44}
\end{eqnarray}%
or%
\begin{eqnarray}
\left\vert \Gamma \right\rangle _{ee}\left\langle \Gamma \right\vert  &=&4%
\genfrac{}{}{0pt}{}{:}{:}\exp \left\{ \frac{\alpha \delta }{\beta
\gamma }\left[ \left( \frac{\sigma _{1}}{\delta
}+\frac{Q_{1}-Q_{2}}{\sqrt{2}}\right) ^{2}+\left( \frac{\sigma
_{2}}{\delta }+\frac{P_{1}+P_{2}}{\sqrt{2}}\right) ^{2}\right]
\right.   \notag \\
&&\left. +\frac{\beta \gamma }{\alpha \delta }\left[ \left( \frac{\tau _{1}}{%
\beta }-\frac{Q_{1}+Q_{2}}{\sqrt{2}}\right) ^{2}+\left( \frac{\tau _{2}}{%
\beta }-\frac{P_{1}-P_{2}}{\sqrt{2}}\right) ^{2}\right] \right\} \genfrac{}{}{0pt}{}{:}{:}%
,  \label{45}
\end{eqnarray}%
which is the Weyl ordered form of $\left\vert \Gamma \right\rangle
_{ee}\left\langle \Gamma \right\vert .$ \allowbreak Noting the difference
between Eq.(\ref{44}) and Eq.(\ref{32}), they are in different operator
ordering. The merit of Weyl ordering lies in the Weyl ordered operators'
invariance under similar transformations, which was proved in Ref.\cite{r19}%
. In addition, it is very convenient for us to obtain the marginal
distributions of $\left\vert \Gamma \right\rangle _{ee}\left\langle \Gamma
\right\vert $ (see the next section).

In Ref. \cite{r20} we have derived the Weyl ordering form of two-mode Wigner
operator $\Delta _{w}\left( \rho ;\varsigma \right) $
\begin{equation}
\Delta _{w}\left( \rho ;\varsigma \right)
=\genfrac{}{}{0pt}{}{:}{:}\delta \left( a_{1}-a_{2}^{\dagger }-\rho
\right) \delta \left( a_{1}^{\dagger }-a_{2}-\rho ^{\ast }\right)
\delta \left( a_{1}+a_{2}^{\dagger }-\varsigma \right) \delta \left(
a_{1}^{\dagger }+a_{2}-\varsigma ^{\ast }\right)
\genfrac{}{}{0pt}{}{:}{:}. \label{46}
\end{equation}%
Eq.(\ref{46}) indicates that the Weyl quantization scheme, for bipartite
entangled operator, is to take the following correspondence,%
\begin{equation}
\rho \rightarrow \left( a_{1}-a_{2}^{\dagger }\right) ,\text{ }\varsigma
\rightarrow \left( a_{1}+a_{2}^{\dagger }\right) ,  \label{47}
\end{equation}%
then the form of Eq.(\ref{44}) indicates that the classical Weyl function
corresponding to $\left\vert \Gamma \right\rangle _{ee}\left\langle \Gamma
\right\vert $ is

\begin{equation}
4\exp \left[ \frac{\alpha \delta }{\beta \gamma }\left\vert \frac{\sigma }{%
\delta }+\rho \right\vert ^{2}+\frac{\gamma \beta }{\alpha \delta }%
\left\vert \frac{\tau }{\beta }-\varsigma \right\vert ^{2}\right] \equiv
h\left( \rho ;\varsigma \right) .  \label{48}
\end{equation}%
Thus the Weyl quantization rule in this case is embodied as%
\begin{eqnarray}
\left\vert \Gamma \right\rangle _{ee}\left\langle \Gamma \right\vert
&=&4\int d^{2}\rho d^{2}\varsigma \genfrac{}{}{0pt}{}{:}{:}\delta
\left( a_{1}-a_{2}^{\dagger }-\rho \right) \delta \left(
a_{1}^{\dagger }-a_{2}-\rho ^{\ast }\right) \delta \left(
a_{1}+a_{2}^{\dagger }-\varsigma
\right)  \notag \\
&&\times \delta \left( a_{1}^{\dagger }+a_{2}-\varsigma ^{\ast }\right)
\genfrac{}{}{0pt}{}{:}{:}\exp \left[ \frac{\alpha \delta }{\beta \gamma }\left\vert \frac{%
\sigma }{\delta }+\rho \right\vert ^{2}+\frac{\gamma \beta }{\alpha \delta }%
\left\vert \frac{\tau }{\beta }-\varsigma \right\vert ^{2}\right]  \notag \\
&=&4\int d^{2}\rho d^{2}\varsigma \Delta _{w}\left( \rho ;\varsigma \right)
\exp \left[ \frac{\alpha \delta }{\beta \gamma }\left\vert \frac{\sigma }{%
\delta }+\rho \right\vert ^{2}+\frac{\gamma \beta }{\alpha \delta }%
\left\vert \frac{\tau }{\beta }-\varsigma \right\vert ^{2}\right] .
\label{49}
\end{eqnarray}%
Using the IWOP technique and Eq.(\ref{31}), in \cite{r21} we have shown that
the normally ordered form of $\Delta _{w}\left( \rho ;\varsigma \right) $ is
\begin{equation}
\Delta _{w}\left( \rho ;\varsigma \right) =\frac{1}{\pi ^{2}}\colon \exp %
\left[ -\left( a_{1}-a_{2}^{\dagger }-\rho \right) \left( a_{1}^{\dagger
}-a_{2}-\rho ^{\ast }\right) -\left( a_{1}+a_{2}^{\dagger }-\varsigma
\right) \left( a_{1}^{\dagger }+a_{2}-\varsigma ^{\ast }\right) \right]
\colon .  \label{50}
\end{equation}%
Substituting Eq.(\ref{50}) into Eq.(\ref{49}) yields%
\begin{eqnarray}
\left\vert \Gamma \right\rangle _{ee}\left\langle \Gamma \right\vert
&=&4\int \frac{d^{2}\rho d^{2}\varsigma }{\pi ^{2}}\colon \exp \left\{
-\left( a_{1}-a_{2}^{\dagger }-\rho \right) \left( a_{1}^{\dagger
}-a_{2}-\rho ^{\ast }\right) +\frac{\alpha \delta }{\beta \gamma }\left\vert
\frac{\sigma }{\delta }+\rho \right\vert ^{2}\right.  \notag \\
&&\left. -\left( a_{1}+a_{2}^{\dagger }-\varsigma \right) \left(
a_{1}^{\dagger }+a_{2}-\varsigma ^{\ast }\right) +\frac{\gamma \beta }{%
\alpha \delta }\left\vert \frac{\tau }{\beta }-\varsigma \right\vert
^{2}\right\} \colon  \notag \\
&=&-4\alpha \beta \gamma \delta \colon \exp \left\{ \frac{\alpha }{\delta }%
\left[ \sigma +\delta \left( a_{1}-a_{2}^{\dagger }\right) \right] \left[
\sigma ^{\ast }+\delta \left( a_{1}^{\dagger }-a_{2}\right) \right] \right.
\notag \\
&&\left. -\frac{\gamma }{\beta }\left[ \tau -\beta \left( a_{2}^{\dagger
}+a_{1}\right) \right] \left[ \tau ^{\ast }-\beta \left( a_{1}^{\dagger
}+a_{2}\right) \right] \right\} \colon ,  \label{51}
\end{eqnarray}%
which confirms Eq.(\ref{32}). In particular, when $\beta =-\delta =1,$ and $%
\alpha =\frac{\kappa }{1+\kappa },$ $\gamma =\frac{1}{1+\kappa },$ Eq.(\ref%
{49}) becomes%
\begin{equation}
\left\vert \Gamma \right\rangle _{ee}\left\langle \Gamma \right\vert
\rightarrow 4\int d^{2}\rho d^{2}\varsigma \Delta _{w}\left( \rho ,\varsigma
\right) \exp \left[ \allowbreak -\kappa \left\vert \rho -\sigma \right\vert
^{2}-\frac{1}{\kappa }\left\vert \varsigma -\tau \right\vert ^{2}\right] ,
\label{52}
\end{equation}%
which is the generalization of single-mode Husimi operator \cite{r22,r23}.

\section{Marginal distributions of $\left\vert \Gamma \right\rangle
_{ee}\left\langle \Gamma \right\vert $}

As mentioned above, based on the Weyl ordered form Eq.(\ref{45}) it is very
convenient for us to obtain the marginal distributions of $\left\vert \Gamma
\right\rangle _{ee}\left\langle \Gamma \right\vert $ ,%
\begin{equation}
\int_{-\infty }^{\infty }\frac{d^{2}\sigma }{\pi }\left\vert \Gamma
\right\rangle _{ee}\left\langle \Gamma \right\vert =-\frac{4\beta
\gamma \delta }{\alpha }\genfrac{}{}{0pt}{}{:}{:}\exp \left\{
\frac{\beta \gamma }{\alpha \delta }\left[ \left( \frac{\tau
_{1}}{\beta }-\frac{Q_{1}+Q_{2}}{\sqrt{2}}\right)
^{2}+\left( \frac{\tau _{2}}{\beta }-\frac{P_{1}-P_{2}}{\sqrt{2}}\right) ^{2}%
\right] \right\} \genfrac{}{}{0pt}{}{:}{:}.  \label{53}
\end{equation}%
Noting $\left[ Q_{1}+Q_{2},P_{1}-P_{2}\right] =0,$ there is no operator
ordering problem involved in Eq.(\ref{53}), so the symbol $\genfrac{}{}{0pt}{}{:}{:}%
\genfrac{}{}{0pt}{}{:}{:}$ in Eq.(\ref{53}) can be neglected,
\begin{equation}
\int_{-\infty }^{\infty }\frac{d^{2}\sigma }{\pi }\left\vert \Gamma
\right\rangle _{ee}\left\langle \Gamma \right\vert =-\frac{4\beta \gamma
\delta }{\alpha }\exp \left\{ \frac{\beta \gamma }{\alpha \delta }\left[
\left( \frac{\tau _{1}}{\beta }-\frac{Q_{1}+Q_{2}}{\sqrt{2}}\right)
^{2}+\left( \frac{\tau _{2}}{\beta }-\frac{P_{1}-P_{2}}{\sqrt{2}}\right) ^{2}%
\right] \right\} .  \label{54}
\end{equation}%
The completeness relation of $|\xi \rangle $ expressed in Eq.(\ref{14}) is%
\begin{eqnarray}
\int \frac{d^{2}\xi }{\pi }\left\vert \xi \right\rangle \left\langle \xi
\right\vert &=&1,\text{ }d^{2}\xi =d\xi _{1}d\xi _{2},  \label{e15} \\
\left\langle \xi ^{\prime }\right. \left\vert \xi \right\rangle &=&\pi
\delta \left( \xi ^{\prime }-\xi \right) \delta \left( \xi ^{\prime \ast
}-\xi ^{\ast }\right) ,
\end{eqnarray}%
we see that the marginal distribution of function $|_{e}\left\langle \Gamma
\right. \left\vert \Psi \right\rangle |^{2}$ in \textquotedblleft $\xi $%
-direction\textquotedblright\ is given by%
\begin{eqnarray}
\left\langle \Psi \right\vert \int_{-\infty }^{\infty }\frac{d^{2}\sigma }{%
\pi }\left\vert \Gamma \right\rangle _{ee}\left\langle \Gamma \right.
\left\vert \Psi \right\rangle &=&\left\langle \Psi \right\vert \int \frac{%
d^{2}\xi }{\pi }\left\vert \xi \right\rangle \left\langle \xi \right\vert
\int_{-\infty }^{\infty }\frac{d^{2}\sigma }{\pi }\left\vert \Gamma
\right\rangle _{ee}\left\langle \Gamma \right. \int \frac{d^{2}\xi ^{\prime }%
}{\pi }\left\vert \xi ^{\prime }\right\rangle \left\langle \xi ^{\prime
}\right. \left\vert \Psi \right\rangle  \notag \\
&=&-\frac{4\beta \gamma \delta }{\alpha }\int_{-\infty }^{\infty }\frac{%
d^{2}\xi }{\pi }\left\vert \Psi \left( \xi \right) \right\vert ^{2}\exp %
\left[ \frac{\beta \gamma }{\alpha \delta }\left\vert \frac{\tau }{\beta }%
-\xi \right\vert ^{2}\right] ,  \label{55}
\end{eqnarray}%
which is a Gaussian-broadened version of quantal distribution $\left\vert
\Psi \left( \xi \right) \right\vert ^{2}$ (measuring two particles' relative
momentum and center-of-mass coordinate)$.$ Similarly, we can obtain another
marginal distribution by performing the integral $d^{2}\tau $ over $%
\left\vert \Gamma \right\rangle _{ee}\left\langle \Gamma \right\vert ,$%
\begin{equation}
\int_{-\infty }^{\infty }\frac{d^{2}\tau }{\pi }\left\vert \Gamma
\right\rangle _{ee}\left\langle \Gamma \right\vert =-\frac{4\alpha \beta
\delta }{\gamma }\exp \left\{ \frac{\alpha \delta }{\beta \gamma }\left[
\left( \frac{\sigma _{1}}{\delta }+\frac{Q_{1}-Q_{2}}{\sqrt{2}}\right)
^{2}+\left( \frac{\sigma _{2}}{\delta }+\frac{P_{1}+P_{2}}{\sqrt{2}}\right)
^{2}\right] \right\} .  \label{56}
\end{equation}%
By using the completeness relation of $|\eta \rangle $,%
\begin{eqnarray}
\int \frac{d^{2}\eta }{\pi }|\eta \rangle \langle \eta | &=&1,\ \ d^{2}\eta
=d\eta _{1}d\eta _{2},  \label{7} \\
\left\langle \eta ^{\prime }\right. \left\vert \eta \right\rangle &=&\pi
\delta \left( \eta ^{\prime }-\eta \xi \right) \delta \left( \eta ^{\prime
\ast }-\eta ^{\ast }\right) ,
\end{eqnarray}%
we see that the other marginal distribution of $|_{e}\left\langle \Gamma
\right. \left\vert \Psi \right\rangle |^{2}$ in \textquotedblleft $\eta $%
-direction\textquotedblright\ is
\begin{equation}
\left\langle \Psi \right\vert \int_{-\infty }^{\infty }\frac{d^{2}\tau }{\pi
}\left\vert \Gamma \right\rangle _{ee}\left\langle \Gamma \right. \left\vert
\Psi \right\rangle =-\frac{4\alpha \beta \delta }{\gamma }\int_{-\infty
}^{\infty }\frac{d^{2}\eta }{\pi }\left\vert \Psi \left( \eta \right)
\right\vert ^{2}\exp \left\{ \frac{\alpha \delta }{\beta \gamma }\left\vert
\frac{\sigma }{\delta }+\eta \right\vert ^{2}\right\} ,  \label{57}
\end{equation}%
a Gaussian-broadened version of quantal distribution $\left\vert \Psi \left(
\eta \right) \right\vert ^{2}$ (measuring two particles' relative coordinate
and total momentum), Eqs.(\ref{55}) and (\ref{57}) describe the relationship
between wave functions in the $_{e}\left\langle \Gamma \right\vert $
representation and those in EPR entangled state $\left\vert \xi
\right\rangle $ ($\left\vert \eta \right\rangle $) representation,
respectively. Note that $|\eta \rangle $ and $\left\vert \xi \right\rangle $
are related to each other by%
\begin{equation}
\left\langle \xi \right. \left\vert \eta \right\rangle =\frac{1}{2}\exp
\left( \frac{\xi ^{\ast }\eta -\xi \eta ^{\ast }}{2}\right) .  \label{15}
\end{equation}

\section{Minimum uncertainty relation for $\left\vert \Gamma \right\rangle
_{e}$}

From the marginal distributions of $\left\vert \Gamma \right\rangle
_{ee}\left\langle \Gamma \right\vert $ we have seen that its phase space
representation involves both the center-of mass (relative) coordinate and
the relative (total) momentum. In order to see clearly how the state $%
\left\vert \Gamma \right\rangle _{e}$ obeys uncertainty relation, we
introduce two pairs of quadrature phase amplitudes for two-mode field:

\begin{equation}
Q_{\pm }\equiv \frac{Q_{1}\pm Q_{2}}{\sqrt{2}},\text{ }P_{\pm }\equiv \frac{%
P_{1}\pm P_{2}}{\sqrt{2}},\text{ }\left[ Q_{\pm },P_{\pm }\right] =i.
\label{58}
\end{equation}%
In similar to deriving Eq.(\ref{36}), using Eqs. (\ref{8}), (\ref{14}) and (%
\ref{20}), we calculate the overlap between $\left\langle \eta \right\vert $
and $\left\vert \Gamma \right\rangle _{e},$
\begin{equation}
\left\langle \eta \right. \left\vert \Gamma \right\rangle _{e}=\sqrt{-\frac{%
\alpha \delta }{\beta \gamma }}\exp \left\{ \frac{\alpha \delta }{2\beta
\gamma }\left\vert \frac{\sigma }{\delta }+\eta \right\vert ^{2}+\frac{1}{%
2\beta }\left[ \tau \left( \eta ^{\ast }-\alpha \sigma ^{\ast }\right) -\tau
^{\ast }\left( \eta -\alpha \sigma \right) \right] \right\} ,  \label{38}
\end{equation}%
and the overlap between $\left\langle \xi \right\vert $ and $\left\vert
\Gamma \right\rangle _{e},$
\begin{equation}
\left\langle \xi \right. \left\vert \Gamma \right\rangle _{e}=\sqrt{-\frac{%
\beta \gamma }{\alpha \delta }}\exp \left\{ \allowbreak \frac{\beta \gamma }{%
2\alpha \delta }\left\vert \frac{\tau }{\beta }-\xi \right\vert ^{2}-\frac{1%
}{2\delta }\left[ \sigma \left( \xi ^{\ast }-\gamma \tau ^{\ast }\right)
-\sigma ^{\ast }\left( \xi -\gamma \tau \right) \right] \right\} .
\label{39}
\end{equation}%
Then employing the completeness relation of $\left\vert \eta \right\rangle $
and Eq.(\ref{38}), we evaluate%
\begin{eqnarray}
\left\langle Q_{-}\right\rangle &=&\int \frac{d^{2}\eta }{\pi }\eta
_{1}\left\vert \left\langle \eta \right. \left\vert \Gamma \right\rangle
_{e}\right\vert ^{2}=-\frac{\sigma _{1}}{\delta },  \notag \\
\left\langle Q_{-}^{2}\right\rangle &=&\int \frac{d^{2}\eta }{\pi }\eta
_{1}^{2}\left\vert \left\langle \eta \right. \left\vert \Gamma \right\rangle
_{e}\right\vert ^{2}=\frac{\sigma _{1}^{2}}{\delta ^{2}}-\frac{\beta \gamma
}{2\alpha \delta },  \label{59}
\end{eqnarray}%
and
\begin{equation}
\left\langle P_{-}\right\rangle =\frac{\tau _{2}}{\beta },\text{ }%
\left\langle P_{-}^{2}\right\rangle =\frac{\tau _{2}^{2}}{\beta ^{2}}-\frac{%
\alpha \delta }{2\beta \gamma }.  \label{60}
\end{equation}%
It then follows
\begin{eqnarray}
\left\langle \Delta Q_{-}^{2}\right\rangle &=&\left\langle
Q_{-}^{2}\right\rangle -\left\langle Q_{-}\right\rangle ^{2}=-\frac{\beta
\gamma }{2\alpha \delta },  \notag \\
\left\langle \Delta P_{-}^{2}\right\rangle &=&\left\langle
P_{-}^{2}\right\rangle -\left\langle P_{-}\right\rangle ^{2}=-\frac{\alpha
\delta }{2\beta \gamma },  \label{61}
\end{eqnarray}%
and%
\begin{equation}
\sqrt{\left\langle \Delta Q_{-}^{2}\right\rangle \left\langle \Delta
P_{-}^{2}\right\rangle }=\frac{1}{2}.  \label{62}
\end{equation}%
In a similar way, using (\ref{39}) we can derive
\begin{eqnarray}
\left\langle Q_{+}\right\rangle &=&\frac{\sigma _{2}}{\delta },\text{ }%
\left\langle Q_{+}^{2}\right\rangle =\frac{\sigma _{2}^{2}}{\delta ^{2}}-%
\frac{\beta \gamma }{2\alpha \delta },  \notag \\
\left\langle P_{+}\right\rangle &=&\frac{\tau _{1}}{\beta },\text{ }%
\left\langle P_{+}^{2}\right\rangle =\frac{\tau _{1}^{2}}{\beta ^{2}}-\frac{%
\alpha \delta }{2\beta \gamma },  \label{63}
\end{eqnarray}%
which also leads to%
\begin{equation}
\sqrt{\left\langle \Delta Q_{+}^{2}\right\rangle \left\langle \Delta
P_{+}^{2}\right\rangle }=\frac{1}{2}.  \label{64}
\end{equation}%
Eqs. (\ref{61})-(\ref{64}) show that $\left\vert \Gamma \right\rangle _{e}$
is a minimum uncertainty state for the two pairs of quadrature operators.

\section{The Wigner function of $\left\vert \Gamma \right\rangle _{e}$}

For a bipartite system, the two-mode Wigner operator in entangled state $%
\left\vert \eta \right\rangle $ representation is expressed in Eq.(\ref{4}),
so the Wigner function $W\left( \rho ,\varsigma \right) $ of $\left\vert
\Gamma \right\rangle _{e}$ is given by
\begin{equation}
W\left( \rho ,\varsigma \right) =Tr\left[ \left\vert \Gamma \right\rangle
_{ee}\left\langle \Gamma \right\vert \Delta _{w}\left( \rho ,\varsigma
\right) \right] =\int \frac{d^{2}\eta }{\pi ^{3}}\left. _{e}\left\langle
\Gamma \right. \left\vert \rho -\eta \right\rangle \right. \left\langle \rho
+\eta \right\vert \left. \Gamma \right\rangle _{e}e^{\eta \varsigma ^{\ast
}-\varsigma \eta ^{\ast }}.  \label{65}
\end{equation}%
Substituting Eq.(\ref{38}) into Eq.(\ref{65}) and using the formula in Eq.(%
\ref{33}), we obtain
\begin{eqnarray}
W\left( \rho ,\varsigma \right)  &=&\int \frac{d^{2}\eta }{\pi ^{3}}\left.
_{e}\left\langle \Gamma \right. \left\vert \rho -\eta \right\rangle \right.
\left\langle \rho +\eta \right\vert \left. \Gamma \right\rangle _{e}e^{\eta
\varsigma ^{\ast }-\varsigma \eta ^{\ast }}  \notag \\
&=&-\frac{\alpha \delta }{\beta \gamma }\int \frac{d^{2}\eta }{\pi ^{3}}\exp
\left\{ \frac{\alpha \delta }{\beta \gamma }\left\vert \eta \right\vert
^{2}+\left( \varsigma ^{\ast }-\frac{\tau ^{\ast }}{\beta }\right) \eta
+\left( \frac{\tau }{\beta }-\varsigma \right) \eta ^{\ast }\right.   \notag
\\
&&\left. +\frac{\alpha }{2\beta \gamma \delta }\left( \allowbreak
2\left\vert \sigma \right\vert ^{2}+2\delta ^{2}\allowbreak \left\vert \rho
\right\vert ^{2}+2\delta \left( \sigma \rho ^{\ast }+\rho \sigma ^{\ast
}\right) \allowbreak \right) \right\}   \notag \\
&=&\frac{1}{\pi ^{2}}\exp \left[ \frac{\alpha \delta }{\beta \gamma }%
\left\vert \frac{\sigma }{\delta }+\rho \right\vert ^{2}+\frac{\gamma \beta
}{\alpha \delta }\left\vert \frac{\tau }{\beta }-\varsigma \right\vert ^{2}%
\right] .  \label{66}
\end{eqnarray}%
From Eq.(\ref{66}) one can see that the Wigner function of $\left\vert
\Gamma \right\rangle _{e}$ possesses the well-behaved feature in the sense
that its marginal distribution in \textquotedblleft $\sigma $%
-direction\textquotedblright\ is a general Gaussian form $\exp \left\{
\allowbreak \frac{\alpha \delta }{\beta \gamma }\left\vert \frac{\sigma }{%
\delta }+\rho \right\vert ^{2}\right\} $, while its marginal distribution in
\textquotedblleft $\tau $-direction\textquotedblright\ is $\exp \left\{
\frac{\beta \gamma }{\alpha \delta }\left\vert \frac{\tau }{\beta }%
-\varsigma \right\vert ^{2}\right\} .$ When $\beta \gamma +\alpha \delta =0\
$and $\beta =-\delta =1,$ Eq.(\ref{66}) reduces to%
\begin{equation}
W\left( \rho ,\varsigma \right) =\frac{1}{\pi ^{2}}\exp \left( -\left\vert
\sigma +\rho \right\vert ^{2}-\left\vert \tau -\varsigma \right\vert
^{2}\right) ,  \label{67}
\end{equation}%
which is just the Wigner function of two-mode canonical coherent state.

Before ending this work, we mention that in 1990 Torres-Vega and Frederick
introduced the state $\left\vert \Gamma \right\rangle $ which satisfies \cite%
{r30}
\begin{eqnarray}
\left\langle \Gamma \right\vert Q_{1} &=&\left( \alpha q+i\beta \frac{%
\partial }{\partial p}\right) \left\langle \Gamma \right\vert ,  \label{69}
\\
\left\langle \Gamma \right\vert P_{1} &=&\left( \gamma p+i\delta \frac{%
\partial }{\partial q}\right) \left\langle \Gamma \right\vert .  \label{70}
\end{eqnarray}%
these two equations well satisfy the correspondence between the classical
and quantum Liouville equations in single-mode case and have many
applications in chemical physics and quantum chemistry \cite{r30}-\cite{r34}%
. Recently $\left\vert \Gamma \right\rangle $ has been identified as a
coherent squeezed state \cite{r29}%
\begin{equation}
\left\vert \Gamma \right\rangle \equiv \left( 2\sqrt{-\alpha \beta \gamma
\delta }\right) ^{1/2}\exp \left[ \frac{\alpha q^{2}}{2\delta }-\frac{\gamma
p^{2}}{2\beta }+\sqrt{2}\left( \alpha q+i\gamma p\right) a_{1}^{\dagger }+%
\frac{\beta \gamma +\alpha \delta }{2}a_{1}^{\dagger 2}\right] \left\vert
0\right\rangle _{1}.  \label{68}
\end{equation}%
The $\left\vert \Gamma \right\rangle _{e}$ in this paper is the non-trivial
generalization of $\left\vert \Gamma \right\rangle $.

In summary,{\small \ }based on the conception of quantum entanglement of
Einstein-Podolsky-Rosen, we have introduced the entangled state $\left\vert
\Gamma \right\rangle _{e}$ for constructing generalized phase space
representation, which possesses well-behaved properties. The set of $%
\left\vert \Gamma \right\rangle _{e}$ make up a complete and
non-orthogonal representation, so it may have new applications, for
examples: 1) It can be chosen as a good representation for solving
dynamic problems for some
Hamiltonians which include explicitly the function of quadrature operators $%
Q_{\pm },$ and/or $P_{\pm };$ 2) $\left\vert \Gamma \right\rangle
_{ee}\left\langle \Gamma \right\vert $ may be considered as a
generalized Wigner operator, since from Eq. (\ref{49}) we see that
it is expressed as smoothing out the usual Wigner operator by
averaging over a \textquotedblleft course graining" function $\exp
\left[ \frac{\alpha \delta }{\beta \gamma }\left\vert \frac{\sigma
}{\delta }+\rho \right\vert
^{2}+\frac{\gamma \beta }{\alpha \delta }\left\vert \frac{\tau }{\beta }%
-\varsigma \right\vert ^{2}\right] ,$ and the corresponding
generalized Wigner function is positive definite. 3) The representation $%
\left\vert \Gamma \right\rangle _{e}$ may be used to analyze
entanglement degree for some entangled states. \textbf{4) } The
$\left\vert \Gamma \right\rangle _{e}$ state can be taken as a
quantum channel for quantum teleportation, such channel may make the
teleportation fidelity flexible, since it involves adjustable
parameters $\alpha ,\beta ,\gamma ,\delta .$ We hope these
applications could be studied in the near future.

\textbf{\ }

\textbf{ACKNOWLEDGEMENTS}

The work is supported by the National Natural Science Foundation of
China under grant no. 10775097.

\textbf{APPENDIX Derivation of Eq.(\ref{20})}

Here we show how to derive Eq.(\ref{20}). In the $|\eta \rangle $
representation we have
\begin{align}
\frac{Q_{1}+Q_{2}}{\sqrt{2}}|\eta \rangle & =-i\frac{\partial }{\partial
\eta _{2}}|\eta \rangle =\left( \frac{\partial }{\partial \eta }-\frac{%
\partial }{\partial \eta ^{\ast }}\right) |\eta \rangle ,  \tag{A1} \\
\frac{P_{1}-P_{2}}{\sqrt{2}}|\eta \rangle & =i\frac{\partial }{\partial \eta
_{1}}|\eta \rangle =i\left( \frac{\partial }{\partial \eta }+\frac{\partial
}{\partial \eta ^{\ast }}\right) |\eta \rangle ,  \tag{A2}
\end{align}%
so according to the requirement in Eqs. (\ref{e5})-(\ref{e6}), we see%
\begin{align}
\left[ \frac{\gamma }{2}\left( \tau +\tau ^{\ast }\right) +\delta \left(
\frac{\partial }{\partial \sigma }-\frac{\partial }{\partial \sigma ^{\ast }}%
\right) \right] \left. _{e}\left\langle \Gamma \right\vert \left. \eta
\right\rangle \right. & =\left( \frac{\partial }{\partial \eta }-\frac{%
\partial }{\partial \eta ^{\ast }}\right) \left. _{e}\left\langle \Gamma
\right\vert \left. \eta \right\rangle \right. ,  \tag{A3} \\
\left[ \frac{\alpha }{2i}\left( \sigma -\sigma ^{\ast }\right) -i\beta
\left( \frac{\partial }{\partial \tau }+\frac{\partial }{\partial \tau
^{\ast }}\right) \right] \left. _{e}\left\langle \Gamma \right\vert \left.
\eta \right\rangle \right. & =\frac{\eta -\eta ^{\ast }}{2i}\left.
_{e}\left\langle \Gamma \right\vert \left. \eta \right\rangle \right. ,
\tag{A4}
\end{align}%
and%
\begin{align}
\left[ \frac{\alpha }{2}\left( \sigma +\sigma ^{\ast }\right) -\beta \left(
\frac{\partial }{\partial \tau }-\frac{\partial }{\partial \tau ^{\ast }}%
\right) \right] \left. _{e}\left\langle \Gamma \right\vert \left. \eta
\right\rangle \right. & =\frac{\eta +\eta ^{\ast }}{2}\left.
_{e}\left\langle \Gamma \right\vert \left. \eta \right\rangle \right. ,
\tag{A5} \\
\left[ \frac{\gamma }{2i}\left( \tau -\tau ^{\ast }\right) +i\delta \left(
\frac{\partial }{\partial \sigma }+\frac{\partial }{\partial \sigma ^{\ast }}%
\right) \right] \left. _{e}\left\langle \Gamma \right\vert \left. \eta
\right\rangle \right. & =i\left( \frac{\partial }{\partial \eta }+\frac{%
\partial }{\partial \eta ^{\ast }}\right) \left. _{e}\left\langle \Gamma
\right\vert \left. \eta \right\rangle \right. ,  \tag{A6}
\end{align}%
Combining Eqs.(A3)-(A6) yields
\begin{align}
\left( \alpha \sigma +2\beta \frac{\partial }{\partial \tau ^{\ast }}\right)
\left. _{e}\left\langle \Gamma \right\vert \left. \eta \right\rangle \right.
& =\eta \left. _{e}\left\langle \Gamma \right\vert \left. \eta \right\rangle
\right. ,  \tag{A7} \\
\left( \alpha \sigma ^{\ast }-2\beta \frac{\partial }{\partial \tau }\right)
\left. _{e}\left\langle \Gamma \right\vert \left. \eta \right\rangle \right.
& =\eta ^{\ast }\left. _{e}\left\langle \Gamma \right\vert \left. \eta
\right\rangle \right. ,  \tag{A8} \\
\left( \gamma \tau ^{\ast }+2\delta \frac{\partial }{\partial \sigma }%
\right) \left. _{e}\left\langle \Gamma \right\vert \left. \eta \right\rangle
\right. & =2\frac{\partial }{\partial \eta }\left. _{e}\left\langle \Gamma
\right\vert \left. \eta \right\rangle \right.  \tag{A9} \\
\left( \gamma \tau -2\delta \frac{\partial }{\partial \sigma ^{\ast }}%
\right) \left. _{e}\left\langle \Gamma \right\vert \left. \eta \right\rangle
\right. & =-2\frac{\partial }{\partial \eta ^{\ast }}\left. _{e}\left\langle
\Gamma \right\vert \left. \eta \right\rangle \right. .  \tag{A10}
\end{align}%
The solution to Eqs.(A7)-(A10) is
\begin{equation}
_{e}\left\langle \Gamma \right\vert \left. \eta \right\rangle =C\exp \left\{
\frac{\alpha \delta }{2\beta \gamma }\left\vert \frac{\sigma }{\delta }+\eta
\right\vert ^{2}+\frac{1}{2\beta }\left[ \tau ^{\ast }\left( \eta -\alpha
\sigma \right) -\tau \left( \eta ^{\ast }-\alpha \sigma ^{\ast }\right) %
\right] \right\} ,  \tag{A11}
\end{equation}%
where $C$ is the normalization constant determined by $_{e}\left\langle
\Gamma \right\vert \left. \Gamma \right\rangle _{e}=1.$

Using the completeness relation of EPR entangled state (\ref{7}) and the
integral formula in Eq.(\ref{33}), we obtain%
\begin{align}
_{e}\left\langle \Gamma \right\vert & =\int \frac{d^{2}\eta }{\pi }%
_{e}\left\langle \Gamma \right\vert \left. \eta \right\rangle \left\langle
\eta \right\vert  \notag \\
& =C\left\langle 00\right\vert \int \frac{d^{2}\eta }{\pi }\exp \left[ -%
\frac{1}{2}|\eta |^{2}+\eta ^{\ast }a_{1}-\eta a_{2}+a_{1}a_{2}\right]
\notag \\
& \times \exp \left\{ \frac{\alpha }{2\beta \gamma \delta }\left\vert \sigma
+\delta \eta \right\vert ^{2}+\frac{1}{2\beta }\left[ \tau ^{\ast }\left(
\eta -\alpha \sigma \right) -\tau \left( \eta ^{\ast }-\alpha \sigma ^{\ast
}\right) \right] \right\}  \notag \\
& =\left\langle 00\right\vert C\exp \left[ \frac{\alpha \left\vert \sigma
\right\vert ^{2}}{2\delta }-\frac{\gamma \left\vert \tau \right\vert ^{2}}{%
2\beta }+\left( \alpha \sigma ^{\ast }+\gamma \tau ^{\ast }\right)
a_{1}+\left( \gamma \tau -\alpha \sigma \right) a_{2}-\left( \beta \gamma
+\alpha \delta \right) a_{1}a_{2}\right] ,  \tag{A12}
\end{align}%
which is just Eq.(\ref{20}) when $C$ is taken as $C=2\sqrt{-\alpha \beta
\gamma \delta }.$

\end{document}